\documentclass[aps,prl,twocolumn,superscriptaddress]{revtex4-2}

\usepackage{graphicx}
\usepackage{amsmath}

\begin{document}

%\preprint{}

\title{
	\large Measuring the photoelectron emission delay in the molecular frame
}

%\author{}
%\thanks{}
%\affiliation{}

\author{Jonas~Rist}
\email{rist@atom.uni-frankfurt.de} 
\author{Kim~Klyssek} 
\affiliation{Institut f\"ur Kernphysik, J. W. Goethe-Universit\"at, Max-von-Laue-Str. 1, D-60438 Frankfurt, Germany}

\author{Nikolay~M.~Novikovskiy}
\affiliation{Institut f\"{u}r Physik und CINSaT, Universit\"{a}t Kassel, Heinrich-Plett-Strasse~40, D-34132 Kassel, Germany}
\affiliation{Institute of Physics, Southern Federal University, 344090 Rostov-on-Don, Russia}

\author{Max~Kircher} 
\author{Isabel~Vela-P\'{e}rez} 
\author{Daniel~Trabert} 
\author{Sven~Grundmann} 
\author{Dimitrios~Tsitsonis} 
\author{Juliane~Siebert} 
\author{Angelina~Geyer} 
\author{Niklas~Melzer} 
\author{Christian~Schwarz} 
\author{Nils~Anders} 
\author{Leon~Kaiser} 
\author{Kilian~Fehre} 
\author{Alexander~Hartung} 
\author{Sebastian~Eckart} 
\author{Lothar~Ph.~H.~Schmidt} 
\author{Markus~S.~Sch\"offler} 
\affiliation{Institut f\"ur Kernphysik, J. W. Goethe-Universit\"at, Max-von-Laue-Str. 1, D-60438 Frankfurt, Germany}

\author{Vernon~T.~Davis}
\author{Joshua~B.~Williams}
\affiliation{Department of Physics, University of Nevada, Reno, Reno, NV 89557, USA}

\author{Florian~Trinter} 
\affiliation{FS-PETRA-S, Deutsches Elektronen-Synchrotron DESY, Notkestr. 85, D-22607 Hamburg, Germany} 
\affiliation{Molecular Physics, Fritz-Haber-Institut der Max-Planck-Gesellschaft, Faradayweg 4-6, D-14195 Berlin, Germany}

\author{Reinhard~D\"orner} 
\affiliation{Institut f\"ur Kernphysik, J. W. Goethe-Universit\"at, Max-von-Laue-Str. 1, D-60438 Frankfurt, Germany}

\author{Philipp~V.~Demekhin}
\email{demekhin@physik.uni-kassel.de} 
\affiliation{Institut f\"{u}r Physik und CINSaT, Universit\"{a}t Kassel, Heinrich-Plett-Strasse~40, D-34132 Kassel, Germany}

\author{Till~Jahnke}
\email{jahnke@atom.uni-frankfurt.de} 
\affiliation{Institut f\"ur Kernphysik, J. W. Goethe-Universit\"at, Max-von-Laue-Str. 1, D-60438 Frankfurt, Germany}


\date{\today}

\begin{abstract}
If matter absorbs a photon of sufficient energy it emits an electron.  The question of the duration of the emission process has intrigued scientists for decades.  With the advent of attosecond metrology, experiments addressing such ultrashort intervals became possible.  While these types of studies require attosecond experimental precision, we present here a novel measurement approach that avoids those experimental difficulties.  We instead extract the emission delay from the interference pattern generated as the emitted photoelectron is diffracted by the parent ion's potential.  Targeting core electrons in CO, we measured a 2d map of photoelectron emission delays in the molecular frame over a wide range of electron energies.  The measured emission times depend drastically on the emission direction and exhibit characteristic changes along the shape resonance of the molecule.  Our approach can be routinely extended to other electron orbitals and more complex molecules.
\end{abstract}

\keywords{Atomic and Molecular Physics} 

\maketitle

\section{I. Introduction}
\begin{widetext}

\begin{figure}[ht]
	\includegraphics[width=14.5cm]{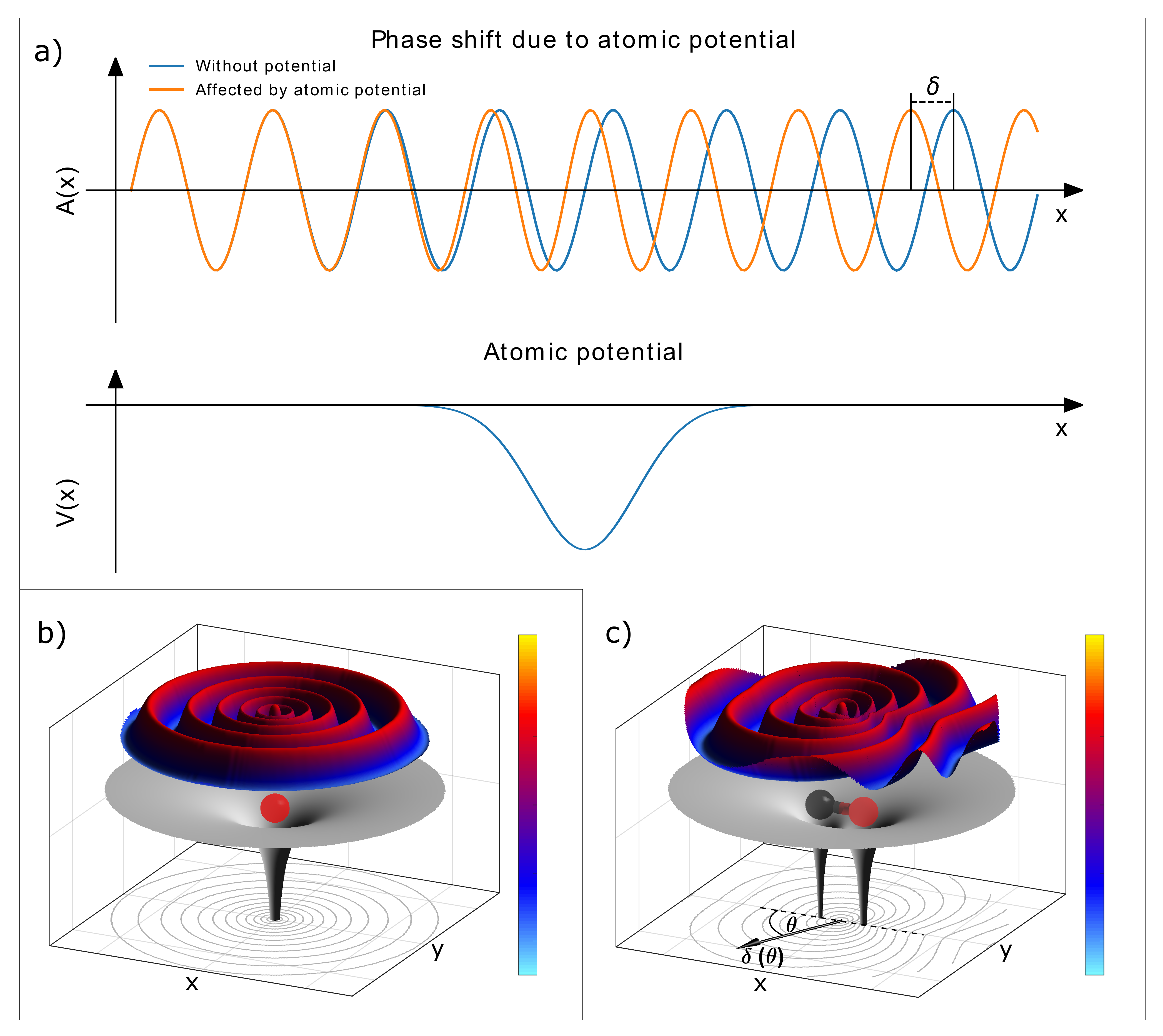}

	\caption{\small{\textbf{Color) Concept of the Wigner phase.} \textbf{(a)} A plane wave with amplitude $A(x)$ passes a potential $V(x)$ from left to right. After passage, it acquires an additional phase $\delta$ due to the modulation of its frequency by the potential. \textbf{(b)} Illustration of an electron wave emitted from within a single atomic potential. \textbf{(c)} The same as (b) but with an additional neighbouring atom resulting in a more complex molecular potential. The photoelectron wave is subject to a phase shift which depends (due to the anisotropy of the potential) on the emission direction with respect to the molecular axis (e.g., $\delta_O \neq \delta_C$, or more generally $\delta$ depends on the emission angle with respect to the molecular axis $\theta$). The sketch depicts as an example the two-well potential of a carbon monoxide molecule.}}
	\label{fig:intro}
\end{figure}

\end{widetext}

The photoelectric effect is one of the most fundamental processes used for probing atoms, molecules and condensed matter. It has been the subject of research for more than a century and most of its aspects are considered well-understood. The basic question of whether the emitted electron appears in the continuum instantaneously or after a short delay has been under investigation for decades. This question, however, needed to be translated into the language of wave mechanics of quantum objects. The translation of the classical concept of a time delay into quantum mechanical wave formalism was first accomplished seventy years ago by Eisenbud and Wigner  (and later Smith) for scattering processes in a series of pioneering theoretical works \cite{ Eisenbud.1948, Wigner.1955, Smith.1960}. Their findings paved the way for the understanding of the concept of a possible photoemission delay. The emitted photoelectron wave is subject to a phase shift induced by the potential of the ionized atom or molecule. This phase shift, as compared to the phase of a wave emerging from a potential-free region, has been termed the \textit{Wigner phase}. The concept of the Wigner phase is depicted in Fig. \ref{fig:intro}. Upon encountering a potential, a plane wave $\Phi_{i}(x,t)=e^{ikx - i \omega t}$ changes its frequency $\omega$ and as a consequence, after interacting with the potential, the phase of the plane wave is shifted by a \textit{scattering phase} $\delta$: $\Phi_{f}(x,t)=e^{ikx - i \omega t + \delta}$ (Fig.~\ref{fig:intro}(a)). The photoeffect mimics that behavior as the photoelectron emerges from within the ion's potential, adding a corresponding \textit{half-scattering} phase (Fig.~\ref{fig:intro}(b)). In the case of molecular photoionization, the situation becomes more complex, as the potential is anisotropic. The Wigner phase acquired in such cases depends on the emission direction of the photoelectron with respect to the molecular axis (Fig.~\ref{fig:intro}(c)). Thus, the phase shift is a particularly sensitive, purely quantum probe of even the most subtle of features of the molecular potential. The photoelectron emission time (often referred to as \textit{Wigner time delay}) is given by the derivative of the electron's phase with respect to the electron's kinetic energy $\varepsilon$ and is in the attosecond regime.

In the aftermath of the initial theoretical investigations, it took several decades before photoemission delays could be addressed in experiments. Several femtosecond-laser-related techniques have been developed during the last 20 years which give experimental access to such atomic time scales. Pioneering work by Schultze \textit{et al.} \cite{Schultze.2010} employed an IR-laser field-streaking approach to measure these ultrashort times. A broadband higher-harmonic attosecond light pulse was used to ionize neon atoms and a superimposed, phase-locked strong IR pulse altered the photoelectron kinetic energy depending on the electron emission time with respect to the IR pulse. From these measurements, the authors concluded that Ne($2p$) electrons are emitted with an additional delay of $21\pm5$~as (attoseconds) compared to those liberated from the Ne($2s$)-shell. This first-of-a-kind work triggered strong theoretical efforts to reproduce the emission delay in the modeling of the process, which mostly yielded much shorter emission delays of only a few attoseconds (see e.g., \cite{Pazourek.2015} for a review). It has been pointed out since then that the streaking laser field alters the electron emission substantially and needs to be incorporated in the models used to extract the actual \textit{naturally occurring} Wigner delay. More recently, the subject of photoionization of Ne has been revisited in a follow-up experiment by Isinger and coworkers \cite{Isinger.2017}. The findings made there suggest that the initial measurement by Schultze \textit{et al.} was (in addition) contaminated by Ne-satellite states with different Wigner delays, which were finally energetically resolved in the recent investigation.

A very detailed theoretical study by Hockett and coworkers targeted Wigner delays in small molecules \cite{Hockett.2016}. Examining the emission delay in CO and N$_2$ molecules, this work provided the first fully three-dimensional Wigner delay maps in the molecular frame and showed the dependence of the emission delay on the electron kinetic energy, the molecular orientation with respect to the light polarization, and the molecular symmetry. Very recently, pioneering work on molecules in the gas phase reported \textit{stereo Wigner delays}, i.e., the difference between emission times along the direction of the carbon and the oxygen atoms of a CO molecule \cite{Vos.2018}. By using a RABBIT scheme (Reconstruction of Attosecond Beating By Interference of Two-photon Transitions) \cite{Paul.2001} and employing a COLTRIMS (Cold Target Recoil Ion Momentum Spectroscopy) reaction microscope \cite{Ullrich.2003}, the authors resolved the phase-beating of IR-induced sidebands in the electron spectrum. From these the stereo Wigner delays for the $\Sigma$- and $\Pi$-orientations of the molecule were obtained for different kinetic energy releases and electron energies. Further work on chiral molecules using the same technique showed that the Wigner delays are enantio-sensitive \cite{Beaulieu.2017} and studies of photoionization time delays in H$_2$ showed a dependence on the dissociation process \cite{Cattaneo.2018}.

\section{II. Experimental Method}
The experimental approach we present in this letter is complementary to the experimental techniques employed so far, which all rely on attosecond light pulses for ionization and on a modification of the emission process by a superimposed strong laser pulse. Our approach is related to the scheme of so-called \textit{complete experiments} \cite{Bederson.1969,Cherepkov.1983}. An emitted electron (wave packet) $\Psi_\varepsilon$ can be written as a coherent superposition of partial waves $\Psi_\varepsilon = \sum a_{\varepsilon \ell m}Y_{\ell m}$ with angular momentum quantum number $\ell $ and magnetic quantum number $m$. Accordingly, if at a given photoelectron energy $\varepsilon$, the complex amplitudes $a_{\varepsilon\ell  m}$ of each partial wave $Y_{\ell m}$ are extracted from an experiment, full information on the emitted electron and the emission process (including the angular-dependent phase) can be retrieved. Furthermore, by scanning the energy of the photon employed for the ionization, the change of the phase as a function of the electron kinetic energy also becomes accessible. This energy derivative is exactly the emission-angle ($\theta,\phi$) and electron-energy ($\varepsilon$) -dependent Wigner delay $t_w$ for which we are looking:

\begin{equation}
	t_w(\varepsilon,\theta,\phi) = \hbar \,\frac{d}{d \varepsilon}\left\{\arg\left[\Psi_\varepsilon(\theta,\phi)\right]\right\}
	\label{WigDel}
\end{equation}

A natural approach in extracting the amplitudes and phases of partial waves contributing to an emitted photoelectron signal is to examine the so-called molecular-frame photoelectron angular distributions (MFPADs). This is because the same scattering of the electron inside the molecular potential that leads to angle-dependent Wigner phases also yields a complex electron diffraction pattern which can be recognized in the emission-direction distribution of the electron with respect to the molecular axis. Photoelectron diffraction imaging \cite{Woodruff.2008} relies, for example, on this effect and employs the measured diffraction pattern to gather insight into the molecular geometry \cite{Wolter.2016}. 
The MFPAD of an electron of a given kinetic energy represents the modulus squared of the coherent superposition of its partial waves:

\begin{equation}
	\frac{d\sigma_\varepsilon}{d\Omega} \sim \left| \Psi_\varepsilon(\theta,\phi) \right|^2 %\sim \left| \sum_{\ell m} a_{\varepsilon \ell m} \,Y_{\ell m}(\theta,\phi) \right|^2
	\label{eq:MFPAD}
\end{equation}

For small molecules, such angular emission distributions can nowadays be obtained routinely using synchrotron light sources for the ionization of distinct molecular orbitals resulting in photoelectrons of a well-defined kinetic energy. 

\section{III. Experimental Setup}

In the present case the experiment was performed at beamline U49/2-PGM-1 of the BESSY II synchrotron \cite{Kachel.2016}. We employed a COLTRIMS reaction microscope \cite{Ullrich.2003} in order to measure the momenta of the photoelectrons and the C$^+$ and O$^+$ ion pairs generated after K-shell ionization and subsequent Auger decay in coincidence. In brief, in the COLTRIMS apparatus, a supersonic gas jet of CO molecules is intersected with the synchrotron photon beam. Charged particles created within the interaction volume by photoionization are guided by homogeneous electric and magnetic fields to two time- and position-sensitive microchannel plate detectors with delay-line position readouts \cite{Jagutzki.2002}. In this experiment, the ion arm consisted of a $5$~cm long acceleration region. The electron arm of the COLTRIMS analyzer incorporated a Wiley-McLaren time-focusing scheme \cite{Wiley.1955} with a $6$-cm acceleration region followed by a $12$-cm field-free drift region. The electric field was set to $13$~V/cm. A superimposed homogeneous magnetic field of $4.3$~Gauss confined electrons up to a kinetic energy of $22$~eV within the spectrometer volume. By measuring the positions of impact and the flight times of all particles in coincidence, the initial momentum vectors are deduced. The MFPADs are then obtained by measuring the emission direction of the C$^+$ and O$^+$ ions, which are generated in a Coulomb explosion after the photoionization process and the subsequent Auger decay. It is known that the ions fragment along the initial molecular axis for an ion kinetic energy release larger than $10.2$~eV \cite{Weber.2001}. We measured the photoelectron momenta in coincidence and, in that way, obtained the relative emission angle, i.e., the emission angle in the molecular frame. The beamline energy was scanned in steps of $50$~meV, and the beamline exit slit was set to $150~${\textmu}m corresponding to a photon energy resolution of approximately $150$~meV.\\

\section{IV. Theoretical Model}

The total amplitude for the emission of a photoelectron with energy $\varepsilon$ in the direction $(\theta,\phi)$ with respect to the axis of a diatomic molecule, which forms the Euler angle $\beta$ with the direction of linear polarization of the ionizing radiation, reads:

\begin{equation}
	\label{eq:ampl}
	T (\varepsilon,\beta,\theta,\phi)=\sum_{\ell m k}(-i)^\ell \, \mathcal{D}^1_{k0}(\beta)\, A_{\varepsilon \ell m k} \,Y_{\ell m}(\theta,\phi).
\end{equation}
Here, $\mathcal{D}^1_{k0}(\beta)$ are the rotation matrices (the remaining orientation Euler angles other than $\beta$ are irrelevant) and $A_{\varepsilon \ell m k}=\langle \Psi_{\varepsilon \ell m} \vert d^1_k \vert \Psi_0\rangle$ are the dipole transition amplitudes for the emission of the partial photoelectron waves  \cite{Cherepkov.1981} with angular momentum quantum numbers $\ell$ and $m$ via the absorption of a photon of polarization $k$, all together given in the frame of a molecule. The amplitudes $A_{\varepsilon \ell m k}$ were computed by the stationary Single Center (SC) method and code \cite{Demekhin.2007,Demekhin.2011,Galitskiy.2015} which provides an accurate theoretical description of the angle-resolved photoemission spectra. The calculations were performed in the frozen-core and relaxed-core Hartree-Fock approximations. The SC expansion of all occupied orbitals of CO was restricted to partial harmonics of $\ell_c \leq 99$, and for the photoelectron in the continuum, to $\ell_\varepsilon \leq 49$. The total transition amplitude in Eq.~(\ref{eq:ampl}) provides access to the MFPADs and Wigner delays via:

\begin{equation}
	\begin{aligned}
		\label{eq:MFPADSdelay}
		\frac{d\sigma}{d\Omega}(\varepsilon,\beta,\theta,\phi)&=\vert T (\varepsilon,\beta,\theta,\phi)\vert^2~~~\\
		\mathrm{and}~~~~t_w(\varepsilon,\beta,\theta,\phi) &= \hbar \, \frac{d}{d \varepsilon}\left\{\arg\left[T (\varepsilon,\beta,\theta,\phi)\right]\right\}.
	\end{aligned}
\end{equation}
The latter derivative was evaluated numerically using energy steps of 100~meV.\\

\section{V. Data Analysis}

In the case that the CO molecules are oriented along the light polarization vector, a considerable simplification of Eq.~(\ref{eq:ampl}) occurs. In particular, the orientation angle $\beta=0$, and the summation over the polarization index reduces to the value $k=0$ (with the respective rotation matrices $\mathcal{D}^1_{00}(0)=1$ and $\mathcal{D}^1_{\pm10}(0)=0$). As a consequence, only $\sigma$-partial waves with $m=0$ contribute to the C $1s$-photoionization channel. Thus, the respective MFPADs can be approximated as:

\begin{equation}
	\frac{d\sigma}{d\Omega} (\varepsilon,\theta) = \left|\sum_{\ell} a_{\varepsilon\ell}\,Y_{\ell 0}(\theta)\right|^2 ~~~\mathrm{with}~~~a_{\varepsilon\ell}=(-i)^\ell \, A_{\varepsilon \ell 0 0} .
	\label{eq:FitEqS}
\end{equation}

In principle, the experimental MFPADs representing the $\Sigma$-channel are obtained by selecting the subset of molecules that are aligned in parallel to the light polarization axis from the whole data set recorded for randomly oriented molecules. However, in order to maximize the statistics of the measured data set we adopted in part the so-called F-function formalism \cite{Lucchese.2002}. The electron angular distribution $I(\beta, \theta, \phi)$ after photoionization can be fully described in terms of the following $F_{LN}$ functions by:

\begin{equation}
	\begin{aligned}
		I(\beta, \theta, \phi)=&F_{00}(\theta) + F_{20}(\theta)P^0_2(\cos(\beta)) \\
		&+F_{21}(\theta)P^1_2(\cos(\beta))\cos(\phi) \\
		&+F_{22}(\theta)P^2_2(\cos(\beta))\cos(2\phi)
	\end{aligned}
\end{equation}

Where, again, $\beta$ is the angle between the molecular axis and the polarization axis and $\theta$ and $\phi$ are the polar and azimuthal angles of the outgoing photoelectron. Setting $\beta = 0$ while integrating over $\phi$ due to the rotational symmetry of the process is equivalent to selecting only the $\Sigma$-orientation. The corresponding $\theta$-dependant distribution is given as:

\begin{equation}
	I_\sigma(\theta) = 4\pi\bigl(F_{00}(\theta) + F_{20}(\theta)\bigr)
\end{equation}

For an isotropic ionization probability, the same angular distribution is obtained by integrating over different values of $\beta$ and combining them in the following way:

\begin{equation}
	I_\sigma(\theta) = 4A(\theta)+\Bigl(4-3\sqrt{3}\Bigr)B(\theta)
\end{equation}

Where $A(\theta)$ and $B(\theta)$ are defined as:

\begin{equation}
	\begin{aligned}
		&A(\theta):=\int_{-1}^{-\frac{1}{\sqrt{3}}} I(x, \theta)dx + \int_{\frac{1}{\sqrt{3}}}^{1} I(x, \theta)dx \\
		&B(\theta):=\int_{-\frac{1}{\sqrt{3}}}^{\frac{1}{\sqrt{3}}} I(x, \theta)dx
	\end{aligned}
\end{equation}

\noindent
with the substitution $x=\cos(\beta)$ and $I(x,\theta) = \int_0^{2\pi} I(x, \theta, \phi) ~d\phi$.

\vspace{1em}

The advantage is obvious for the experiment because now all recorded events contribute to the MFPAD of the $\Sigma$-state (instead of only a small fraction) which leads to significantly better statistics. 

A similar procedure can be followed for molecules oriented perpendicularly to the light polarization ($\beta = \pi/2$) if only that part of the MFPAD distribution is considered which lies within the plane defined by the molecular axis and the polarisation axis of the light ($\phi = 0$). In this case only $\pi$-partial waves with $m = \pm1$ emitted via absorption of photons with polarization $k=\pm1$ contribute to the ionisation (the respective rotation matrices are equal to $\mathcal{D}^1_{00}(\frac{\pi}{2})=0$ and $\mathcal{D}^1_{\pm10}(\frac{\pi}{2})=\mp\frac{1}{\sqrt{2}}$). Thereby, Eq.~(\ref{eq:ampl}) simplifies to:

\begin{equation}
	\begin{aligned}
		\frac{d\sigma}{d\Omega} (\varepsilon,\theta) &= \left| -\sqrt{2}\sum_{\ell} a_{\varepsilon\ell}  Y_{\ell 1}(\theta) \right|^2 \\
		\mathrm{with}~~~a_{\varepsilon\ell}&=(-i)^\ell \, A_{\varepsilon \ell 1 1} .
		\label{eq:FitEqP}
	\end{aligned}
\end{equation}
Where we used $Y_{l1}=-Y_{l(-1)}$ for $\phi=0$ and $A_{\varepsilon \ell 1 1} = A_{\varepsilon \ell (-1)(-1)}$.
\vspace{1.em}\\
As before, the F-function formalism can be used to enhance the statistics for the case of $\Pi$-orientation by setting $\beta=\pi\text{/}2$ and $\phi = 0$:

\begin{equation}
	\begin{aligned}
		I_\pi(\theta) &= I(\pi/2,\theta, 0) = F_{00}(\theta) + \frac{1}{2}F_{20}(\theta) + 3F_{22}(\theta) \\
		&= \frac{1}{4\pi}X(\theta)+\frac{1}{8\pi}Y(\theta) + \frac{3}{16}Z(\theta)
	\end{aligned}
\end{equation}

Where $X(\theta)$, $Y(\theta)$ and $Z(\theta)$ are defined as:

\begin{equation}
	\begin{aligned}
		&X(\theta):=A(\theta)+B(\theta) \\
		&Y(\theta):=3A(\theta)+\Bigl(3-3\sqrt{3}\Bigr)B(\theta) \\
		&Z(\theta):=8\int_{0}^{\frac{1}{4}\pi} I(\theta, \phi)d\phi
	\end{aligned}
\end{equation}
with $x=\cos(\beta)$ and $I(\theta, \phi) = \int_{-1}^{1} I(x, \theta, \phi)dx$.

\vspace{1em}

We used the Minuit2 package of the ROOT data-analysis framework \cite{Brun.1997} to fit Eq.~(\ref{eq:FitEqS}) and Eq.~(\ref{eq:FitEqP}) to our MFPADs obtained by the F-function formalism in steps of $100$~meV in the electron energy. For every fit the events within a range of $\pm 1~eV$ were integrated. The experimentally-derived distributions are reproduced adequately when restricting the summation to spherical harmonics of $\ell \leq 4$ (as has already been previously demonstrated \cite{Jahnke.2002}). The main challenge of such multi-parameter fitting of the real-valued MFPAD in Eq.~(\ref{eq:FitEqS}) and Eq.~(\ref{eq:FitEqP}) with the complex amplitudes $a_{\varepsilon\ell}$ is a lack of uniqueness. In particular, all complex amplitudes can be extracted up to one common global phase, which is, in turn, energy-dependent. As a consequence, the fitting procedure can yield random jumps of this global phase as a function of energy. We employed the following solution to this problem. First, for each energy, the isotropic contribution to the total amplitude $a_{\varepsilon 0}\,Y_{0 m}$ was set to be real. As a result, all fitted amplitudes were determined up to the unknown energy-dependent global phase $\delta_0(\varepsilon)$ of the amplitude $a_{\varepsilon 0}$. Second, our theoretical calculations showed that the phase $\arg\left[\Psi_\varepsilon(\theta)\right]$, as a function of the emission angle $\theta$, depicted a monotone behavior. To fit the first data set (of lowest electron energy), we therefore initialized the fitting algorithm with random parameters and selected a result that fulfilled that monotonicity condition. Using the fitted result obtained in that step as an input to the fitting of the next adjacent energy step, we obtained results that are consistent with our theoretical calculations. This procedure yielded reliable and stable results for many different sets of random initial parameters. We obtained the Wigner delays by using adjacent energy steps to numerically evaluate the derivative (given here, for example for $\Sigma$-orientation)

\begin{equation}
	\begin{split}
		t_w^{\sigma}(\varepsilon,\theta) &= \hbar \,\frac{\arg\left[\sum_{\ell} a_{\varepsilon\ell}Y_{\ell 0}(\theta)\right]- \arg\left[\sum_{\ell} a_{\varepsilon^\prime\ell}Y_{\ell 0}(\theta)\right] }{\varepsilon - \varepsilon^\prime }
		\label{WigEq}
	\end{split}
\end{equation}

\noindent
and found that an energy step of $\varepsilon - \varepsilon^\prime = 1$~eV yielded the most stable delays while still reproducing the details of the theoretical predictions. It should be stressed, however, that Eq.~(\ref{WigEq}) yields experimental Wigner delays on a relative scale, i.e., up to an unknown (but isotropic) energy-dependent delay, $t^0_w(\varepsilon)=\hbar \frac{d\delta_0(\varepsilon)}{d\varepsilon}$, provided by the amplitude $a_{\varepsilon \ell}$. This missing isotropic delay cannot be determined from the experiment but can be fixed by calibration to the theory. Here,  for each photoelectron energy, we set the angle-averaged experimental Wigner delay to the respective theoretical value.

\section{VI. Results}

\begin{widetext}

\begin{figure}[ht]
	\centering
	\includegraphics[width=16.5cm]{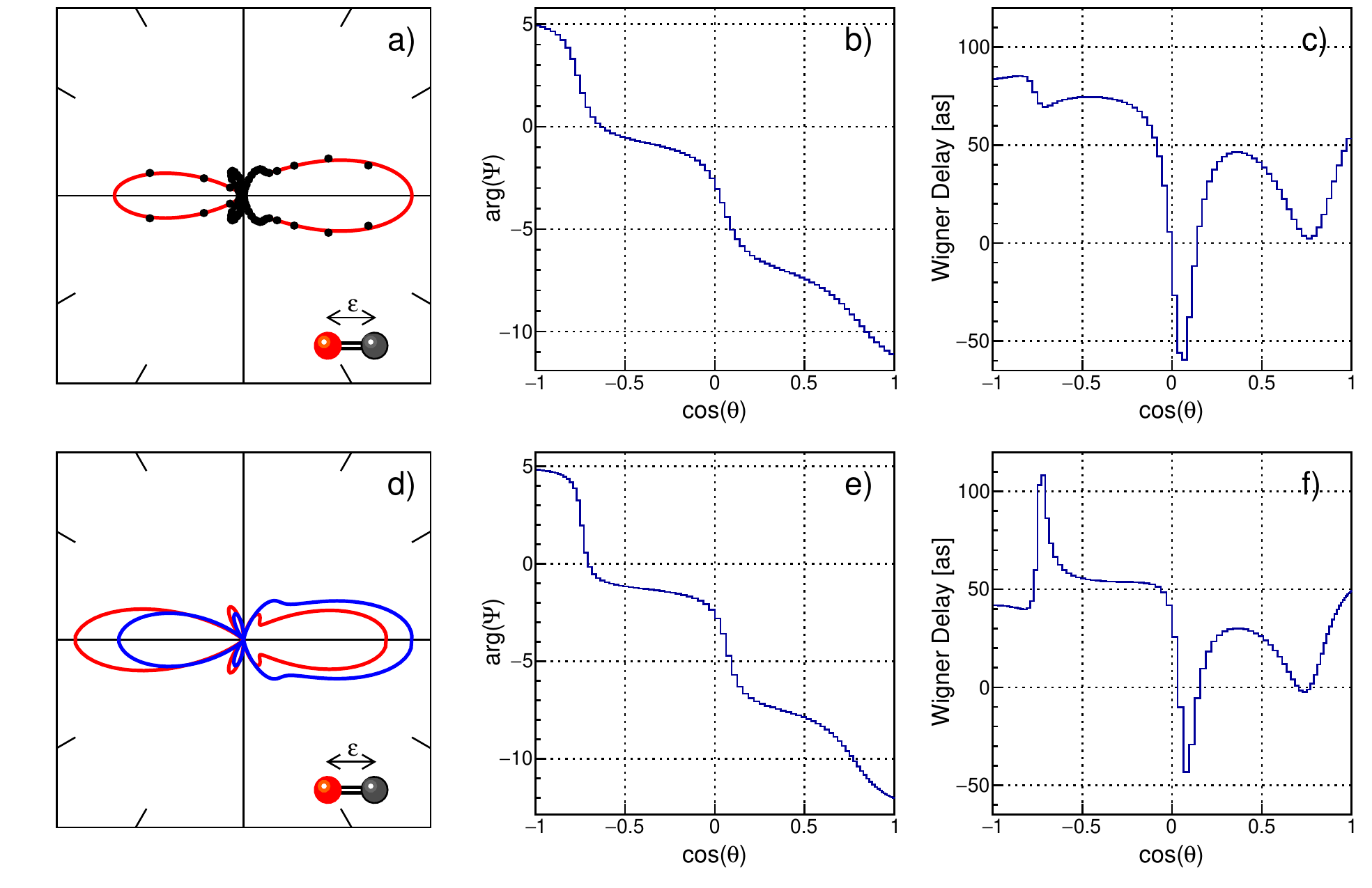}
	\caption{\small{\textbf{Color) Molecular-frame photoelectron angular distribution and extracted information for $\boldsymbol{\varepsilon=18.8}$~eV.} \textbf{(a)} Example of the emission pattern of a core electron emitted from the carbon atom of a CO molecule. The electron has been ionized by linearly-polarized synchrotron light. The molecule is oriented horizontally and parallel to the polarization axis of the photons with the emitting carbon atom pointing to the right (as depicted by the inset). The rich angular features are caused by the scattering of the emerging electron wave by the molecular potential. The statistical error bars of the data points are smaller than the markers. \textbf{(b)} The extracted phase $\arg\left[\Psi_\varepsilon(\theta)\right]$ and \textbf{(c)} molecular-frame  Wigner delay. The oxygen atom is located at $\cos(\theta)=-1$, the carbon atom at $\cos(\theta)=1$. \textbf{(d)--(f)} Corresponding results obtained from our theoretical modeling of the photoemission process. The two lines in panel (d) belong to the modeling within the relaxed-core (red) and the frozen-core (blue) Hartree-Fock approximation. The phase and the Wigner delay shown in the panels (e) and (f) correspond to the relaxed-core Hartree-Fock approximation.}}
	\label{fig:MFPAD}
\end{figure}

\end{widetext}

An example of such an angular distribution obtained from ionizing carbon monoxide molecules is given in Fig. \ref{fig:MFPAD}(a). The red line is a fit using Eq.~\ref{eq:FitEqS} to the measured data points employing partial waves up to $\ell=4$. From the fitting results the phase, $\arg\left[\Psi_\varepsilon(\theta)\right]$, is calculated for each photoelectron energy and according to Eq.~\ref{WigDel} the corresponding molecular-frame Wigner delays are obtained. Both are shown in Fig. \ref{fig:MFPAD}(b) and \ref{fig:MFPAD}(c), respectively. 

To confirm our experimental findings, we performed a modeling of the photoemission process using Hartree-Fock wave functions. The corresponding results are depicted in the lower panels of Fig. \ref{fig:MFPAD} and show good agreement with the experimental findings. The two distributions shown in Fig.~\ref{fig:MFPAD}(d) have been computed using a relaxed-core (red line) and a frozen-core (blue line) Hartree-Fock approach. The experimental data can be considered to lie in between the two theoretical computations, slightly favoring the latter model. The measured and computed phase $\arg\left[\Psi_\varepsilon(\theta)\right]$  decreases monotonically as one moves from the direction in which the oxygen atom points towards the direction in which the carbon atom points. The molecular-frame derivative of the phase, i.e., the Wigner delay, exhibits more distinct features. The computed results show a sharp maximum and a deep minimum. The minimum can be clearly observed in the experimental results as well. The maximum is less pronounced but occurs in the progression towards lower photoelectron kinetic energies (see next paragraph). The Wigner delay varies by several 10s of attoseconds depending on the emission angle of the photoelectron, just as predicted by Hockett \textit{et al.} \cite{Hockett.2016} for the valence ionization of CO.

\begin{widetext}

	\begin{figure}[bt]
		\centering
		\includegraphics[width=13cm]{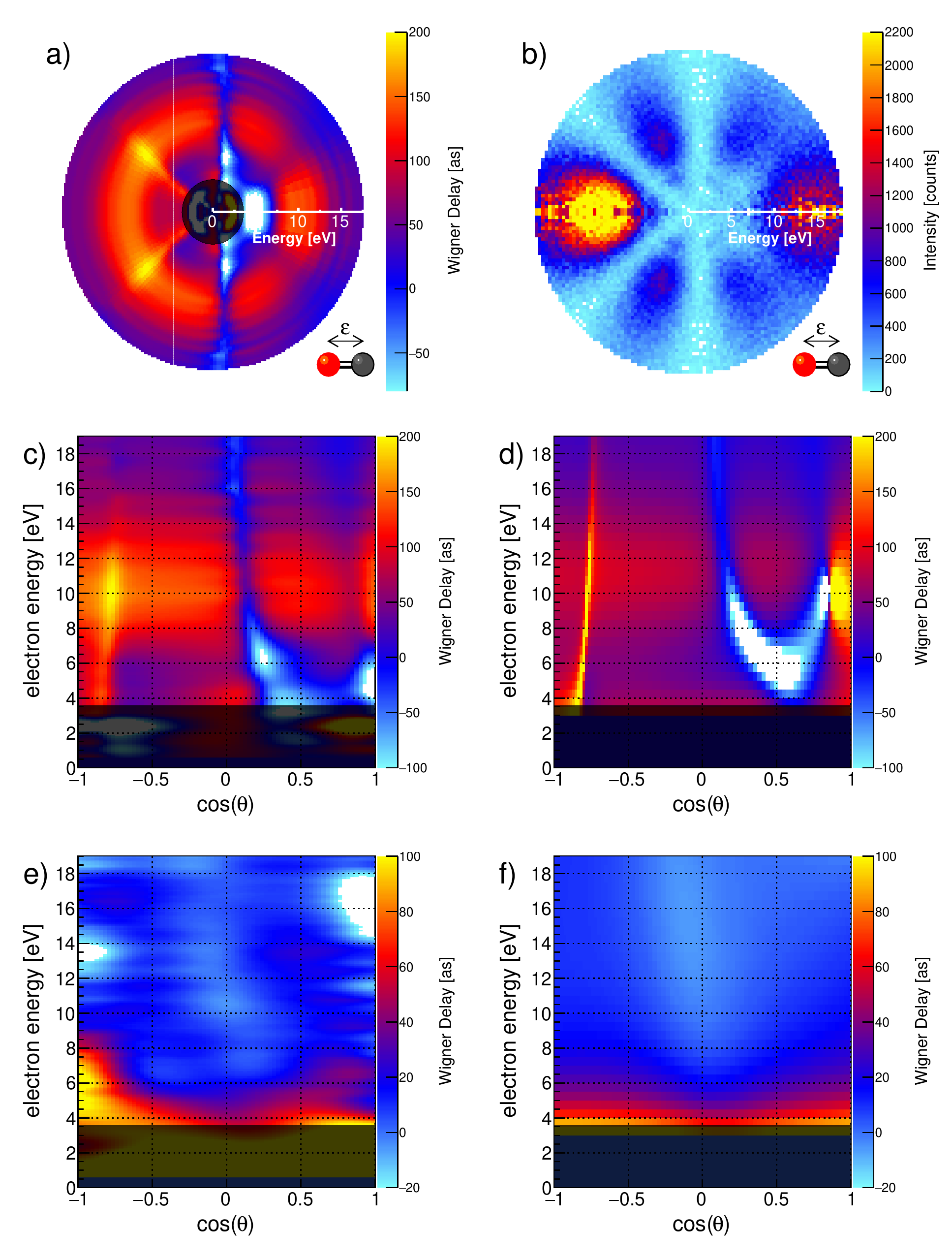}
		\caption{\small{\textbf{Color) Molecular-frame photoelectron angular distributions and Wigner delay maps.} \textbf{(a)--(d)} The molecule is oriented along the light's polarization axis. \textbf{(a)} Angle-dependent Wigner delay for a range of electron kinetic energies. The electron energy is encoded in the distance from the center, while the value in [as] is encoded in the color scale. \textbf{(b)} Polar map of the MFPAD in a corresponding representation. Distinct features in the Wigner delay occur under emission angles which depict minima in the angular emission distribution. \textbf{(c)} The same as in (a) but in a conventional color map representation. \textbf{(d)} Associated Wigner color map obtained from our theoretical modeling. \textbf{(e)} and \textbf{(f)} The same as (c) and (d) but for molecules oriented perpendicularly to the polarization axis of the incoming photons. For the energy region below $3.5$~eV a different theoretical model would be necessary. Therefore, this region is grayed out in (a) and (c)--(f).}}
		\label{fig:MFPAD2D}
	\end{figure}
\end{widetext}

%\begin{widetext}
%	\begin{figure}
%		\begin{minipage}[c]{0.67\textwidth}
%			\includegraphics[width=\textwidth]{Fig3_3x2.pdf}
%		\end{minipage}\hfill
%		\begin{minipage}[c]{0.3\textwidth}
%			\caption{\small{\textbf{Color) Molecular-frame photoelectron angular distributions and Wigner delay maps.} \textbf{(a)--(d)} The molecule is oriented along the light's polarization axis. \textbf{(a)} Angle-dependent Wigner delay for a range of electron kinetic energies. The electron energy is encoded in the distance from the center, while the value in [as] is encoded in the color scale. \textbf{(b)} Polar map of the MFPAD in a corresponding representation. Distinct features in the Wigner delay occur under emission angles which depict minima in the angular emission distribution. \textbf{(c)} The same as in (a) but in a conventional color map representation. \textbf{(d)} Associated Wigner color map obtained from our theoretical modeling. \textbf{(e)} and \textbf{(f)} The same as (c) and (d) but for molecules oriented perpendicularly to the polarization axis of the incoming photons. For the energy region below $3.5$~eV a different theoretical model would be necessary. Therefore, this region is grayed out in (a) and (c)--(f).}} 	
%		\label{fig:MFPAD2D}
%		\end{minipage}
%	\end{figure}
%\end{widetext}

We have performed additional measurements of MFPADs and Wigner delays in the range of the first $20$~eV above the CO carbon K-threshold. Figure \ref{fig:MFPAD2D} compiles the results. The 2D color maps shown in Figs. \ref{fig:MFPAD2D}(a) and \ref{fig:MFPAD2D}(b) depict the Wigner delay as a function of electron kinetic energy and in a polar distribution of intensity, respectively. The electron energy is encoded in the radial distance from the plot's center. Fig.~\ref{fig:MFPAD2D}(a) shows the molecular-frame Wigner delay map in this plotting scheme. The corresponding measured MFPAD is presented in Fig.~\ref{fig:MFPAD2D}(b). This representation of the data shows that distinct features in the Wigner delay occur at the same emission angles as minima in the MFPAD. This behavior is in line with the predictions that drastic changes in the emission delay may occur in case of the destructive interference of partial waves due to two-center or Cohen-Fano interference effects \cite{Ning.2014}.

For comparison, the energy-dependent Wigner delay map resulting from our theoretical modeling and the map of the experimental data are shown in Figs.~\ref{fig:MFPAD2D}(d) and \ref{fig:MFPAD2D}(c) as color maps as functions of the molecular-frame photoelectron emission angle and the electron energy. The molecular axis is aligned along the direction of the light polarization ($\Sigma$-orientation). In addition, we present the corresponding histograms for molecules oriented perpendicularly to the light's polarization axis (i.e., $\Pi$-orientation) in Figs.~\ref{fig:MFPAD2D}(e) and \ref{fig:MFPAD2D}(f). The range of observed Wigner delays is much smaller for the $\Pi$-case. This is expected, as a main feature of the photoemission spectrum in the presented electron energy regime is a $\Sigma$-shape resonance, which is a broad resonance feature appearing only for the $\Sigma$-orientation with a maximum at approx. 8~eV \cite{Shigemasa.1993}. Shape resonances result from a trapping of the emerging electron wave inside a centrifugal barrier which is present for higher-angular-momentum contributions to the emitted electron wave \cite{Piancastelli.1999}. We clearly observe how the Wigner delay range increases as the ionization energy progresses across the resonance in the electron energy range of $5$~eV $\lesssim  E_e \lesssim  12$~eV  in Figs. \ref{fig:MFPAD2D}(c)+\ref{fig:MFPAD2D}(d), while Figs. \ref{fig:MFPAD2D}(e)+\ref{fig:MFPAD2D}(f) do not depict this behavior. Furthermore, the average Wigner delay, when weighted with the angular emission probability, is approximately $130$~as on top of the resonance, which agrees nicely with the resonance's width of approximately $5$~eV to $6$~eV. The theoretical modeling (Fig. \ref{fig:MFPAD2D}(d)) reveals a particularly strong feature at $E_e = 9.1$~eV. The color scale is cropped in order to highlight the other details of the Wigner delay map, but the feature diverges towards negative delays and then rises abruptly to large positive delays within a very small angular range close to $\cos(\theta)=0.8$. We do not observe a corresponding feature in the measured Wigner delay map in Fig.~\ref{fig:MFPAD2D}(c) and attribute the overestimated strength to the fixed equilibrium internuclear distance of the CO molecule employed in our theoretical modeling. We estimate that those close-lying negative and positive delays will compensate for each other if the nuclear motion is included. We notice, however, that in spite of the fixed-nuclei one-particle Hartree-Fock approximation, which is known to suffer accuracy problems where shape resonances are concerned \cite{Dehmer.1975}, the present theory reproduces all the features observed in the experiment. The photoelectron energy region below $3.5$~eV is grayed out in the histograms. It is known that in this energy region, the photoionization is strongly affected by doubly excited states \cite{Cherepkov.2000}, which are not included in our theoretical model. The angular dependence observed in the experiment, however, suggests, that these doubly excited states give rise to particularly interesting modulations in the emission time, which could be the subject of future investigations.

\section{VII. Conclusion}
In summary, we presented a novel experimental approach to extract photoelectron Wigner delays in the molecular frame. Contrary to previous approaches, measuring this attosecond-level observable does not require ultrashort laser pulses or even attosecond pulses to trigger photoemission and our approach measures the native photoionization process of the unperturbed molecule without any need for dressing laser fields. Furthermore, the use of low-bandwidth synchrotron radiation allows for addressing electrons in a wide range of binding energies and distinct orbitals even in larger molecules. For more complex molecules, which are not cylindrically symmetric, complete 3D Wigner delay maps can be extracted. In the future, we will address excitation energy ranges which are dominated by electron correlation effects|as for example those of the aforementioned doubly excited states|in order to explore the role of correlation in the Wigner phase. It has been recently demonstrated that MFPADs can be measured with X-ray free-electron lasers using the same multi-particle coincidence approach employed in the present experiment \cite{Kastirke.2020,Kastirke.2020b}. These measurements suggest that MFPADs (and thus 3D Wigner delay maps) from more complex molecules and processes can be obtained using X-ray free-electron laser sources and more importantly, that the temporal evolution of the MFPADs following a photoreaction will become accessible in the near future.

%\appendix
%\section{Experimental Setuo}

\subsection*{Author contribution}
All authors except for P.V.D. and N.M.N. contributed to the experiment. K.K, J.R. and T.J. performed the data analysis and fitting procedure. P.V.D. and N.M.N. performed the theoretical modeling. All authors contributed to the manuscript.

\subsection*{Autor Information}
The authors declare no competing financial interests. Correspondence and requests for materials should be addressed to J.R. (rist@atom.uni-frankfurt.de) or T.J. (jahnke@atom.uni-frankfurt.de).

\subsection*{Acknowledgments}
\begin{acknowledgments}
	The experimental work was supported by the Deutsche Forschungsgemeinschaft (DFG) and the Bundesministerium f\"ur Bildung und Forschung (BMBF). S.E. and D.T. were supported by DFG Priority Programme ``Quantum Dynamics in Tailored Intense
	Fields'' (Project No. DO 604/29-1). A.H. and K.F. are grateful for the support of the Studienstiftung des deutschen Volkes. T.J., S.G., R.D and M.S.S. acknowledge support from the DFG via Sonderforschungsbereich 1319 (ELCH). The theoretical work was supported by the DFG Project No. DE 2366/1-2. We are very thankful for the outstanding support provided by the BESSY II staff as part of the Helmholtz Zentrum Berlin (HZB), in particular that of Ronny Golnak. T.J. is very grateful for initial discussion on the topic with Danielle Dowek. R.D. and T.J. thank Nikolai Cherepkov and Ricardo  D\'{\i}ez Mui\~no for initial education on fitting partial waves many years ago.
\end{acknowledgments}

\bibliography{ArXiv_CO_Wigner.bib}

\end{document}